\documentclass[10pt,conference]{IEEEtran}
\IEEEoverridecommandlockouts
\usepackage{cite}
\usepackage{amsmath,amssymb,amsfonts}
\usepackage{algorithmic}
\usepackage{graphicx}
\usepackage{textcomp}
\usepackage{xcolor}
\usepackage{url}
\def\BibTeX{{\rm B\kern-.05em{\sc i\kern-.025em b}\kern-.08em
    T\kern-.1667em\lower.7ex\hbox{E}\kern-.125emX}}
\usepackage{tikz}
\usetikzlibrary{shapes.geometric, arrows.meta, positioning, fit, backgrounds, calc}
\makeatletter
\newcommand{\linebreakand}{%
  \end{@IEEEauthorhalign}
  \hfill\mbox{}\par
  \mbox{}\hfill\begin{@IEEEauthorhalign}
}
\makeatother
\begin{document}
\title{From Legacy Documentation to OSCAL: An MCP-Based Agent Pipeline for Threat-Informed Continuous Compliance in Critical Infrastructure}
\author{%
\IEEEauthorblockN{Lea Muth}
\IEEEauthorblockA{Department of Mathematics and Computer Science \\
Freie Universität Berlin\\
Berlin, Germany \\
Lea.Muth@fu-berlin.de}
\and
\IEEEauthorblockN{Marian Margraf}
\IEEEauthorblockA{Department of Mathematics and Computer Science \\
Freie Universität Berlin\\
Berlin, Germany \\
Marian.Margraf@fu-berlin.de}
}

\maketitle

\begingroup
\renewcommand\thefootnote{}
\footnotetext{\textcopyright~2026 IEEE. Personal use of this material is permitted. Permission from IEEE must be obtained for all other uses, in any current or future media, including reprinting/republishing this material for advertising or promotional purposes, creating new collective works, for resale or redistribution to servers or lists, or reuse of any copyrighted component of this work in other works.}
\endgroup

\begin{abstract}
In critical infrastructure, operational technology environments often cannot be actively scanned, and yet active system feedback is needed for risk assessment and compliance. This paper presents a non-invasive, MCP-grounded multi-agent pipeline that converts natural-language system descriptions into source-verified knowledge graph and audit-ready artifacts in the NIST OSCAL format for continuous automated compliance management. The architecture decouples LLM-based reasoning from deterministic knowledge retrieval against authoritative threat-intelligence sources, reducing the risk of fabricated vulnerabilities and hallucinated attack paths.

In an evidence-based synthetic scenario of a water utility, the pipeline achieves 0.90 CVE recall and perfect D3FEND recall. It generates a schema-valid OSCAL System Security Plan and an OSCAL Security Assessment Report. Nevertheless, the core insight is not that grounding via MCP eliminates errors (e.g., hallucinations) entirely from the pipeline, but that it shifts errors into the first phase of asset extraction from the natural language description. Here, a single incorrectly extracted entity can lead to genuine but irrelevant CVEs in subsequent stages of the pipeline, which consumes time and resources. However, it makes the remaining risk visible, verifiable, and suitable for a time-efficient manual review, since the infrastructure (e.g., version numbers, OS, etc.) is typically known.
\end{abstract}


\section{Introduction}
In mid-September 2025, the global cybersecurity landscape underwent a significant tactical escalation. Anthropic’s Threat Intelligence team disclosed a sophisticated AI-orchestrated espionage campaign, presumably executed by the state-sponsored group GTG-1002~\cite{anthropic2025gtg1002}. 
For the first time in documented history, threat actors leveraged an autonomous framework based on Anthropic’s Claude Code and the Model Context Protocol (MCP) to execute complex cyber operations with minimal human intervention (80\%-90\% automated tactical operations), demonstrating that the primary threat of AI lies not in generating novel zero-day exploits, but in accelerating the overall attack lifecycle~\cite{anthropic2025gtg1002}. 
This acceleration poses a severe systemic risk to critical infrastructure, where the defensive posture is often constrained by operational technology. critical infrastructure sectors such as energy and water management rely on legacy Industrial Control Systems (ICS) with life cycles that span decades. Given the fragility of these legacy operational technology environments, standard security measures (e.g., active vulnerability scanning) pose unacceptable availability risks~\cite{stouffer2015}. Consequently, while attackers already use Multi-Agent Systems with MCP, defenders still lack real-time visibility and often rely on fragmented static asset documentation.
%
While AI-powered tools promise to close this gap, they also introduce fundamental reliability problems, as they hallucinate Common Vulnerabilities and Exposures (CVE) identifiers, fabricate Common Vulnerability Scoring System (CVSS) scores, and generate plausible but fictitious attack paths~\cite{syed2025agentic2}. In operational technology, hallucinated attack paths can trigger shutdowns leading to catastrophic scenarios. Without active scanning of the environment, defenders have no way of evaluating the statements made by AI tools.
%
Another challenge is the gap in reliability, as regulatory bodies have tightened compliance requirements, e.g., through NIS-2~\cite{NIS2}, which requires rigorous and well-documented risk management. In September 2025, the German Federal Office for Information Security (BSI) released the new Grundschutz++ requirements as machine-readable Open Security Controls Assessment Language (OSCAL) catalogs~\cite{bsi_sdtb_github}, creating the technical precondition for automated compliance. However, the mere availability of the JSON schema does not bridge this gap, as critical infrastructure operators possess unstructured, natural-language system descriptions (e.g., legacy documentation, operator-provided infrastructure summaries) and not the structured models required for automated AI reasoning.

To address the above challenges, this paper proposes a non-invasive eight-phase pipeline utilizing a Multi-Agent System to transform natural-language system descriptions into audit-ready OSCAL security artifacts. To minimize hallucinations, the pipeline implements knowledge grounding through the MCP standard. Fifteen MCP servers integrate authoritative Cyber Threat Intelligence (CTI) sources into the agent inference process. Through the deterministic and LLM-assisted phases of the pipeline, agents iteratively enrich the Knowledge Graph (KG) with vulnerabilities, urgency ratings, attack paths, and complete taxonomy chains. The resulting KG is converted into NIST OSCAL artifacts, specifically a System Security Plan and a Security Assessment Report, enabling direct integration into audit workflows without requiring invasive active scans. 
Evaluated across five runs against a synthetic evidence-based hybrid IT and operational technology water utility scenario, the framework achieves a stable CVE Recall of $0.90$, a D3FEND Recall of $1.00$, and a Factual Hallucination Rate of $0\%$ for all deterministically sourced KG nodes. An entity extraction evaluation against manually verified ground truth reveals a Semantic Hallucination Rate of 12.5\%, with the resulting Contextual False Positive Rate of 8.5\% quantifying the error propagation from LLM-based entity extraction into the deterministic retrieval pipeline.
The main contributions of this work are:
\begin{itemize}
  \item An MCP-grounded eight-phase pipeline for transforming unstructured critical infrastructure documentation into audit-ready OSCAL artifacts without active scanning, enabling direct integration into audit workflows, shifting failures from distributed LLM hallucination to a localized, human-reviewable control point (Phase 0).
  \item An empirically grounded, domain-specific heuristic for vulnerability prioritization in operational technology that combines CVSS, EPSS, and KEV with asset criticality and network exposure for safety-first triage.
\end{itemize}
\section{Background}
\subsection{The Model Context Protocol}
MCP is an open standard that provides a structured interface between LLM-based applications and external data sources or tools~\cite{mcp_github}. The pipeline uses 15 active MCP servers grouped into six functional categories as shown in Table~\ref{tab:mcp_servers}.
\begin{table}[htbp]
\centering
\caption{Active MCP Servers by Pipeline Category}
\label{tab:mcp_servers}
\scriptsize
\begin{tabular}{|p{1.4cm}|p{2.8cm}|p{2.8cm}|}
\hline
\textbf{Category} & \textbf{Active Servers} & \textbf{Reason} \\
\hline
Vuln.\newline Foundations & NVD, CPE, US-CERT,\newline ICS-CERT, OpenCVE\enspace(5) & CVE/CPE discovery\newline and advisory enrichment \\
\hline
Risk Filters & KEV, EPSS\enspace(2) & Exploitation evidence\newline and probability scoring \\
\hline
Infrastructure\newline Visibility & Shodan\enspace(1) & Passive exposure\newline reconnaissance \\
\hline
Threat\newline Taxonomies & ATT\&CK Enterprise,\newline ATT\&CK ICS, EMB3D,\newline CAPEC, CWE\enspace(5) & Attack pattern and\newline weakness chaining \\
\hline
Resilience\newline \& Defense & D3FEND\enspace(1) & Defensive counter-\newline measure mapping \\
\hline
Lifecycle \&\newline Supply Chain & EOL\enspace(1) & End-of-life and\newline obsolescence detection \\
\hline
\end{tabular}
\end{table}
In security assessment, MCP's main advantage is constrained retrieval, as agents are prevented from hallucinating a CVE or a CVSS score, since they must execute pre-defined functions against validated endpoints. Furthermore, MCP ensures that responses are returned in structured JSON format rather than unstructured text, allowing schema validation before data ingestion into the KG. 

A fundamental limitation of traditional Retrieval-Augmented Generation (RAG) in cybersecurity is its reliance on vector similarity, which does not encode the strict causal and transitive logic required for risk analysis, as contrasted in Table~\ref{tab:rag_vs_mcp}. Robust threat reasoning requires precise, verifiable links rather than semantic proximity alone.
\begin{table}[htbp]
\centering
\caption{Traditional RAG vs. MCP Tool-Use in CTI}
\label{tab:rag_vs_mcp}
\scriptsize
\begin{tabular}{|p{1.8cm}|p{2.8cm}|p{2.8cm}|}
\hline
\textbf{Dimension} & \textbf{RAG} & \textbf{MCP Tool-Use} \\
\hline
Retrieval type & Semantic similarity\newline (vector search) & Deterministic query\newline (structured API call) \\
\hline
Verification & Probabilistic\newline (cosine threshold) & Binary\newline (exists or does not) \\
\hline
Error class & Hallucinated\newline associations & Missing data\newline (API gaps) \\
\hline
Freshness & Static corpus\newline (re-indexing required) & Real-time\newline (live API queries) \\
\hline
operational technology suitability & Low (false positives\newline trigger wrong decisions) & High (verifiable\newline ground truth) \\
\hline
\end{tabular}
\end{table}
\subsection{Threat-Centric Metrics}
\label{sec:vuln_scoring}
The Exploit Prediction Scoring System (EPSS) provides a probability ($0-1$) that a specific vulnerability will be exploited within the next 30 days~\cite{jacobs2021epss}, and the Cybersecurity and Infrastructure Security Agency (CISA) Known Exploited Vulnerabilities (KEV) catalog serves as a confirmation (binary) of active exploitation by threat actors~\cite{cisa2021bod2201}.
This distinction is fundamental to the empirically calibrated heuristic proposed later in this paper, as the CVSS base score defines the potential impact of a vulnerability, while EPSS and KEV define the urgency of the threat. The heuristic combines these dimensions to isolate action-able threats from theoretical noise and ensure that limited defense resources are focused primarily on high-probability attack vectors.
\subsection{The NIST OSCAL Framework}
The OSCAL was designed by NIST to standardize the exchange of security assessment data in machine-readable form including the following three artifacts. The System Security Plan (SSP) describes the target system's boundaries, components, and the implementation status of required security controls. 
The Security Assessment Report (SAR) is a structured assessment report that links specific technical findings (e.g., CVEs, attack paths) back to the control requirements defined within the SSP, thereby identifying gaps between the required and the implemented controls. 
Although OSCAL defines a separate Security Assessment Plan (SAP), our framework encodes the plan directly from the deterministic pipeline configuration (selected MCP sources, assessment thresholds, and taxonomy traversal rules). These deterministic phases act as an implicit, reproducible SAP without any manual assessment plan creation required.
\section{Related Work}
\subsection{MCP-Based Approaches in Security Research}
Prior work can be grouped into security analyses of MCP ecosystems and MCP-enabled security-agent architectures.
Within the MCP ecosystem security literature, the authors of~\cite{hou2025mcp} provide an structured analysis of MCP attack surfaces across the server life cycle and develop a taxonomy of attacker types, threat scenarios, and corresponding safeguards.
In~\cite{hasan2025mcp}, the authors add large-scale empirical evidence from 1,899 open-source MCP servers (343 official, 1,556 community), using static analysis plus an MCP-specific scanner. They report eight MCP-relevant vulnerability categories and quantify prevalence rates, including 7.2\% general vulnerabilities, 5.5\% tool-poisoning exposure, 66\% code smells, and 14.4\% traditional bug patterns.
The authors of~\cite{radosevich2025mcp} demonstrate practical exploitation against mainstream LLMs through unmodified MCP servers and formalize the three attack classes of Malicious Code Execution, Remote Access Control abuse, and Credential Theft. They also released the agentic multi-stage auditing framework MCPSafetyScanner.

Within the field of MCP-Enabled Security Agents, the authors of~\cite{kurniawan2025agcyrag} propose AgCyRAG, a modular multi-agent cybersecurity RAG architecture with role-specialized agents for guardrailing, retrieval, SPARQL-based KG access, and answer refinement. Their SPARQL path uses MCP-mediated tool invocation over the SEPSES cybersecurity KG (CVE, CWE, CPE, CAPEC, ATT\&CK). 
The authors of~\cite{syed2025agentic} combine LLM reasoning, reinforcement learning, and multi-agent coordination for CI/CD supply-chain security in a LangGraph stack, with MCP used as a communication layer for pipeline-integrated actions.
In~\cite{narajala2025enterprise}, the authors analyze enterprise MCP risks via a MAESTRO-based, multi-layer threat-modeling perspective and discuss deployment controls like stronger authentication and monitoring. Closely related hardening mechanisms such as immutable tool and version governance are developed in subsequent ETDI-oriented work~\cite{bhatt2025etdi}.

To the best of our knowledge, no existing work combines MCP as a deterministic CTI interface layer, operational technology/critical infrastructure-specific KG construction with attack-path derivation, and OSCAL-validated audit-ready compliance artifacts. AgCyRAG~\cite{kurniawan2025agcyrag} is architecturally closest but operates as a single-framework RAG system for retrospective analysis rather than proactive multi-phase threat modeling with epistemic guardrails.
\subsection{Risk Prioritization in Operational Technology}
In operational technology and critical infrastructure environments, established vulnerability scores such as CVSS are only of limited use. Primarily as they are modeling IT impacts and effects, whereas in industrial systems, a single vulnerability can cause severe cascade effects. CVSS v3.1 is not reflecting this safety dimension in practice and although CVSS v4.0 introduces the Supplemental Safety value, it does not directly influence the base score~\cite{cvss40_user_guide}. Furthermore, its real-world applicability is currently limited because CVSS 4.0 assessments are only available for a fraction of all CVEs (as of March 2026). For reliable operational technology prioritization, a pure CVSS view is therefore not sufficient.
Alternative frameworks address specific parts of the problem, but they do not replace an overall open, consistent operational technology score. CISA's Stakeholder-Specific Vulnerability Categorization (SSVC) uses a decision tree rather than a numerical value to prioritize, taking factors such as mission impact, automation capabilities, safety impact, and exploitation status into account. In particular, the safety branch explicitly distinguishes between IT and operational technology consequences. The MITRE EMB3D is a valuable index for threat modeling of operational technology devices, but it is also not a universal prioritization index.

Research further demonstrates the necessity for extensions of classical scores for cyber-physical contexts and multi-signal prioritization. The authors of \cite{rvss_score} developed the Robot Vulnerability Scoring System (RVSS), which extends CVSS to include safety and physical impact aspects as well as cascade effects. It has been tested in practice but it is heavily robotics-centric. In their work~\cite{kausar2026cpps} the authors combine multiple threat intelligence signals and show for ICS-relevant CVEs, that linking CVSS, EPSS, KEV, and operational technology characteristics improves prioritization, but with limited transparency and a high need for manual labeling. The authors of \cite{shimizu2025vmchain} validate the effectiveness of CVSS+EPSS+KEV in IT contexts, but do not include operational technology context factors. In \cite{cheimonidis2024dvcc}, the authors demonstrate the advantages of environment-dependent adaptation using the Dynamic Vulnerability Severity Calculator, but do not integrate signals regarding the timeliness of threats. Generally, surveys such as \cite{jiang2025vulnsurvey} confirm the lack of an open context-sensitive operational technology standard.

To counteract the aforementioned problem, we developed the critical infrastructure Relevance Heuristic (CRH), as it is designed as a transparent, numerical, and adjustable index. Methodologically, it adopts the safety-first logic from SSVC, the timeliness dimension from EPSS/KEV, the environmental adaptation approach from DVCC, and the empirically supported multi-signal combination from newer vulnerability management approaches. Unlike black-box machine learning models, the weighting and structure remain transparent and traceable, as CRH explicitly integrates the operational technology/critical infrastructure context (criticality of assets, network exposure/purdue segmentation) in contrast to IT-centric frameworks. The heuristics support automated, threshold-based triage without any loss of transparency.
\subsection{OSCAL in Compliance-as-Code}
The authors of~\cite{oscal_paper} demonstrate the integration of attack graphs into the NIST OSCAL ecosystem. Their framework validates the suitability of OSCAL as a pivot format for compliance automation and employs NIST SP~800-53 controls with DSL-based attack graphs for risk assessment.
But the authors assume the existence of highly structured input data, specifically hardware and software bills of materials and formal DSL models in their work. In legacy operational technology environments, such structured artifacts rarely exist. Our approach addresses the gap between unstructured natural-language system descriptions and OSCAL, functioning as an automated ingestion layer for legacy environments.
Furthermore, the authors aggregate temporal CVSS metrics via BFS traversal on DSL graphs. Our CRH heuristic combines CVSS with EPSS/KEV and operational technology-specific context factors (asset criticality, exposure) for active threat prioritization rather than static severity aggregation.
For verification, they rely on the correlation between BOMs and vulnerability databases. Our approach enforces deterministic API calls against primary sources through MCP, with each KG node being verified individually, providing a verifiable ground truth over hallucinated vulnerability assignment.
\section{Theoretical Framework: Multi-Agent System Architecture and Experimental Setup}
\subsection{Pipeline Phases}
\label{sec:pipeline_phases}
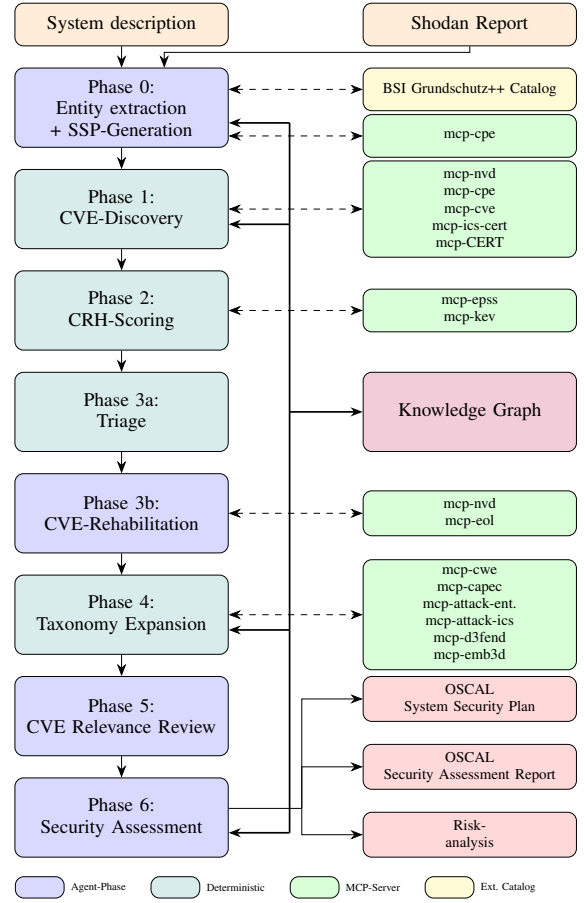
\begin{figure}[htbp]
\centering
\resizebox{0.84\columnwidth}{!}{%
\begin{tikzpicture}[
  node distance=0.35cm and 1.0cm,
  >={Stealth[length=2mm]},
  box/.style={rectangle, draw, rounded corners, minimum width=3.5cm, minimum height=0.7cm,
      align=center, font=\small},
  llm/.style={box, fill=blue!15, minimum height=1.3cm},
  det/.style={box, fill=teal!15, minimum height=1.3cm},
  prog/.style={rectangle, draw, rounded corners, fill=gray!12, minimum width=2.2cm,
      minimum height=0.5cm, align=center, font=\scriptsize},
  mcp/.style={box, fill=green!15, font=\scriptsize},
  ext/.style={box, fill=yellow!20, font=\scriptsize},
  data/.style={box, fill=orange!15},
  output/.style={box, fill=red!15, font=\scriptsize},
  lbl/.style={font=\scriptsize\bfseries, text=gray}
]

\node[data] (input) {System description};
\node[data, right=2.2cm of input] (shodan) {Shodan Report};

\node[llm, below=of input] (phase0) {Phase 0:\\Entity extraction\\+ SSP-Generation};
\node[ext, anchor=north west] (ext0b) at ([xshift=2.2cm]phase0.north east) {BSI Grundschutz++ Catalog};
\node[mcp, below=0.05cm of ext0b] (mcp0b) {mcp-cpe};

\node[det, below=of phase0] (phase2) {Phase 1:\\CVE-Discovery};
\node[mcp, right=2.2cm of phase2] (mcp1) {mcp-nvd\\mcp-cpe\\mcp-cve\\mcp-ics-cert\\mcp-CERT};

\node[det, below=of phase2] (phase3) {Phase 2:\\CRH-Scoring};
\node[mcp, right=2.2cm of phase3] (mcp2) {mcp-epss\\mcp-kev};

\node[det, below=of phase3] (phase4a) {Phase 3a:\\Triage};

\node[llm, below=of phase4a] (phase4b) {Phase 3b:\\CVE-Rehabilitation};

\node[det, below=of phase4b] (phase5) {Phase 4:\\Taxonomy Expansion};
\node[mcp, right=2.2cm of phase5] (mcp4) {mcp-cwe\\mcp-capec\\mcp-attack-ent.\\mcp-attack-ics\\mcp-d3fend\\mcp-emb3d};

\node[llm, below=of phase5] (phase6) {Phase 5:\\CVE Relevance Review};

\node[llm, below=of phase6] (phase7) {Phase 6:\\Security Assessment};

\node[output, anchor=north west] (out1) at ([xshift=2.2cm]phase6.north east) {OSCAL\\System Security Plan};
\node[output, anchor=south west] (out3) at ([xshift=2.2cm]phase7.south east) {Risk-\\analysis};
\node[output] (out2) at ($(out1)!0.5!(out3)$) {OSCAL\\Security Assessment Report};

\node[box, fill=purple!20, minimum height=1.3cm,
  right=2.2cm of phase4a] (kg) {Knowledge Graph};

\node[mcp, right=2.2cm of phase4b] (mcp3b) {mcp-nvd\\mcp-eol};

\draw[->] (input) -- (phase0);
\draw[->] (shodan.south) |- ([yshift=0.25cm, xshift=0.7cm]phase0.north)
-- ([xshift=0.7cm]phase0.north);
\draw[->] (phase0) -- (phase2);
\draw[->] (phase2) -- (phase3);
\draw[->] (phase3) -- (phase4a);
\draw[->] (phase4a) -- (phase4b);
\draw[->] (phase4b) -- (phase5);
\draw[->] (phase5) -- (phase6);
\draw[->] (phase6) -- (phase7);

\coordinate (kgbus) at ([xshift=1.0cm]phase4a.east);
\draw[->, thick] (kgbus |- kg.west) -- (kg.west);
\draw[-, thick] ([yshift=-0.25cm]phase0.east -| kgbus) -- ([yshift=-0.25cm]phase7.east -| kgbus);
\draw[<-, thick] ([yshift=-0.25cm]phase0.east) -- ([yshift=-0.25cm]phase0.east -| kgbus);
\draw[<-, thick] ([yshift=-0.25cm]phase2.east) -- ([yshift=-0.25cm]phase2.east -| kgbus);
\draw[<-, thick] ([yshift=-0.25cm]phase5.east) -- ([yshift=-0.25cm]phase5.east -| kgbus);
\draw[<-, thick] ([yshift=-0.25cm]phase7.east) -- ([yshift=-0.25cm]phase7.east -| kgbus);

\draw[<->, dashed] (phase0.east |- ext0b.center) -- (ext0b.west);
\draw[<->, dashed] (phase0.east |- mcp0b.center) -- (mcp0b.west);
\draw[<->, dashed] (phase2.east) -- (mcp1.west);
\draw[<->, dashed] (phase3.east) -- (mcp2.west);
\draw[<->, dashed] (phase5.east) -- (mcp4.west);
\draw[<->, dashed] (phase4b) -- (mcp3b);

\coordinate (outfork) at ([xshift=1.2cm, yshift=0.15cm]phase7.east);
\draw[-] ([yshift=0.15cm]phase7.east) -- (outfork);
\draw[->] (outfork) |- (out1.west);
\draw[->] (outfork) |- (out2.west);
\draw[->] (outfork) |- (out3.west);

\path let \p1=(phase7.south west), \p2=(out3.south east) in
coordinate (legL) at (\x1, \y1-0.3cm)
coordinate (legR) at (\x2, \y1-0.7cm);
\node[anchor=north west, inner sep=0pt] at (legL) {
  \begin{tikzpicture}[node distance=0.15cm and 0.6cm]
      \node[rectangle, draw, rounded corners, fill=blue!15,
          minimum width=0.8cm, minimum height=0.35cm] (l1) {};
      \node[right=0.1cm of l1, font=\tiny] (t1) {Agent-Phase};
      \node[rectangle, draw, rounded corners, fill=teal!15,
          minimum width=0.8cm, minimum height=0.35cm, right=0.4cm of t1] (l2) {};
      \node[right=0.1cm of l2, font=\tiny] (t2) {Deterministic};
      \node[rectangle, draw, rounded corners, fill=green!15,
          minimum width=0.8cm, minimum height=0.35cm, right=0.4cm of t2] (l3) {};
      \node[right=0.1cm of l3, font=\tiny] (t3) {MCP-Server};
      \node[rectangle, draw, rounded corners, fill=yellow!20,
          minimum width=0.8cm, minimum height=0.35cm, right=0.4cm of t3] (l4) {};
      \node[right=0.1cm of l4, font=\tiny] (t4) {Ext. Catalog};
  \end{tikzpicture}
};

\end{tikzpicture}%
}
\caption{Architecture of the eight-phase pipeline: agent-supported phase (blue), deterministic phase (turquoise), MCP-backed data sources (green), BSI Grundschutz++ (yellow), and the central KG (pink). Output artifacts (red): OSCAL SSP, OSCAL SAR, and CRH-based Risk Assessment.}
\label{fig:architecture}
\end{figure}

The pipeline utilizes a graph-based Multi-Agent System to convert unstructured natural-language system descriptions into machine-readable compliance artifacts in OSCAL format. To adhere to the non-invasive operational constraints of operational technology infrastructures, the system operates strictly on passive input artifacts.
As illustrated in Fig.~\ref{fig:architecture}, the transformation process is orchestrated through an eight-phase pipeline that incrementally enriches the central KG. The architecture enforces a strict separation between probabilistic LLM-driven phases (0, 0b, 3b, 5, 6) and deterministic phases (1, 2, 3a, 4).

The KG is a directed graph with seven node types (Entity type, CVE, CWE, CAPEC, ATT\&CK, and D3FEND) and connected by six typed edges: \textit{has\_vulnerability} (Entity to CVE), \textit{has\_weakness} (CVE to CWE), \textit{exploited\_by} (CWE to CAPEC), \textit{maps\_to\_technique} (CAPEC to ATT\&CK), \textit{exploits\_technique} (CVE to ATT\&CK), and \textit{countered\_by} (ATT\&CK to D3FEND). 
Each pipeline phase appends nodes and edges incrementally, so that after Phase~4 a fully expanded chain reads: Entity$\to$CVE$\to$CWE$\to$CAPEC$\to$ATT\&CK$\to$D3FEND.
\subsubsection{Phase~0 - Entity Extraction \& SSP} Phase 0 initializes the KG by transforming a system description and a Shodan report into a structured asset inventory. The agent semantically decomposes the descriptions into hardware and software components, assigning each a vendor name, version number, topological role (Gateway/Pivot/Target), a type (PLC/HMI/Workstation/\ldots), a criticality (critical/high/medium/low), and an exposure score (internet/DMZ/internal/isolated). 
The SSP is assembled from the previously extracted entities by serializing metadata, component definitions, CPE assignments, and BSI Grundschutz++ control implementation statements into a valid OSCAL JSON, which then serves as the baseline for the dynamic infrastructure KG and all subsequent pipeline phases.
\subsubsection{Phase~1 - CVE-Discovery} Phase 1 uses a multi-stage discovery strategy. It begins by resolving CVE identifiers already present in extracted entities via mcp-nvd. Next, it performs CPE-based retrieval with normalized CPE strings (e.g., 9.2(4) $\to$ 9.2.4). To capture sector-specific findings that may not surface through plain CPE matching, it queries CISA ICS-CERT advisories by manufacturer names. Coverage is then expanded through alias CPE lookups for renamed products, followed by keyword- and concept-based fallback searches for entities without valid CPE matches, enabling the detection of vulnerabilities without standardized identifiers. All discovered CVEs are linked to their corresponding entities in the KG and annotated with strategy-specific confidence scores: 1.0 (CPE match), 0.9 (ICS-CERT match), 0.7 (keyword match), and 0.5 (fallback).
\subsubsection{Phase~2 - CRH-Scoring} In Phase 2, all discovered CVEs are evaluated using the CRH. The CRH operates according to the principle of "safety-first approximation" to resolve the systematic discrepancy between severity levels (CVSS) and operational operational technology risks. Mainly because in critical infrastructures an overlooked critical vulnerability (false negative) has far more devastating consequences than an irrelevant alarm (false positive). The parameters are therefore intentionally conservative in order to strictly prioritize recall over precision. 
The urgency index $U$ is calculated additively according to~\eqref{eq:CRH}, where $U$ denotes the Urgency Index, $S_{\text{base}}$ is the CVSS-derived severity base score, $M_{\text{context}}$ is the asset-context multiplier, and $B_{\text{threat}}$ is an additive threat intelligence bonus.
\begin{equation}
U = (S_{\text{base}} \times M_{\text{context}}) + B_{\text{threat}}
\label{eq:CRH}
\end{equation}
The components are defined as follows:
\begin{equation}
S_{\text{base}} = \max(\text{CVSS}, 5.0)
\label{eq:s}
\end{equation}
The CVSS base score in~\eqref{eq:s} is limited to a minimum of $5.0$ by a Supervisory Control and Data Acquisition (SCADA) floor. This reflects the operational technology-specific character, as a CVE with a low CVSS score affecting a PLC carries far greater risk in critical infrastructure contexts than the score suggests, as physical processes are directly at stake. The threshold of $5.0$ - the midpoint of the CVSS scale (0–10) - lifts low-rated operational technology vulnerabilities into the NVD "Medium" category while preserving full scoring granularity above it. 
The context multiplier $M_{\text{context}}$ is composed of asset criticality $C_{\text{crit}}$ and exposure factor $E_{\text{exp}}$:
\begin{equation}
M_{\text{context}} = C_{\text{crit}} \times E_{\text{exp}}
\label{eq:m}
\end{equation}
The criticality levels $C_{\text{crit}}$ are inspired by the Purdue model~\cite{ackerman2024industrial}, since it ranks operational technology assets according to their proximity to the physical process (Level 0-5), resulting in a criticality classification at the asset level. Critical ($1.5$) addresses assets at Level~0-1 (sensors, PLCs, RTUs, etc.), operational ($1.25$) corresponds to Level~2 (HMI, SCADA servers), and support ($1.0$) forms the baseline for Level~3+ (operation management, DMZ, enterprise~IT). 
The exposure $E_{\text{exp}}$ distinguishes between Internal ($1.0$, accessible internally) and Internet-Exposed ($1.5$, visible through Shodan). The maximum single multiplier is limited to $1.5$ in order to maintain CVSS granularity. A higher multiplier would place almost all critically exposed assets in the highest risk category, counteracting the differentiating purpose of the CVSS baseline value. 
\begin{equation}
B_{\text{threat}} = \text{EPSS} \times 5 + \begin{cases} 5 & \text{if CVE} \in \text{KEV} \\ 0 & \text{else} \end{cases}
\label{eq:b}
\end{equation}
The additive threat bonus $B_{\text{threat}}$ from~\eqref{eq:b} is limited to the range $[0, 10]$, which means that active CTI can significantly influence the triage decision, but cannot determine it alone. Hence, a critical rating always requires a high CVSS base score and/or high context multiplier (critical asset, exposed). Therefore, $B_{\text{threat}}$ sums the scaled EPSS probability ($\text{EPSS} \times 5$) and a binary penalty for KEV entries ($+5$ points). This penalty ensures that KEV-listed CVEs are always significantly elevated regardless of their EPSS value.
The symmetric weighting of both terms reflects two fundamental yet orthogonal sources of information. An asymmetrical distribution would inevitably result in one term dominating the other. 
The additive calculation of $B_{\text{threat}}$, in contrast to the multiplication used in classic risk models (e.g., $risk = threat * vulnerability * impact$), prevents the methodological error of Score Masking. The additive integration guarantees that the potential for physical damage remains visible, regardless of the current statistical exploit probability.
The Urgency Index is limited to an absolute maximum value ($U \le 32.5$) and classified into four Urgency Categories (Critical $\geq 22$, High $\geq 14$, Medium $\geq 7$, Low $< 7$). 
It should be emphasized that the selected constants are not universally optimal but empirically derived values. In the absence of an established gold standard for context-aware operational technology vulnerability prioritization, they serve as a transparent approximation with parameters that are disclosed to enable domain experts to customize them for specific systems.
\subsubsection{Phase~3 - Triage \& Rehabilitation} Phase 3 has two stages. In Phase 3a, CVEs are triaged if they satisfy any of three criteria: an absolute threshold ($U>12$), a role-based threshold for gateway/target assets ($U>5$), or KEV inclusion (unconditional override). High-urgency CVEs are then used to build attack paths across taxonomy chains (gateway $\to$ pivot $\to$ target), while all others move to Phase 3b.
In agent-based Phase 3b, discarded CVEs are re-evaluated against the SSP component inventory (CPEs, vendors, asset types) using enriched CVE metadata and MCP-provided affected-products data for CPE-to-CPE matching instead of text heuristics. CVEs linked to expired end-of-life (EOL) assets are always rehabilitated, as there is no vendor fix, only workarounds or replacement.

\subsubsection{Phase~4 - Taxonomy Expansion} The top 25 high-urgency CVEs are expanded along complete taxonomy chain. Each chain node is inserted into the KG and linked using typed edges. 
A maximum of 20 taxonomy chains are retained per CVE. For operational technology assets, ICS-specific ATT\&CK techniques are prioritized and mapped to threats for embedded devices using mcp-emb3d.

\subsubsection{Phase~5 - Relevance Review} In Phase~5, the agent analyzes the remaining CVEs for additional relevance and retroactively classifies a maximum of 5 additional CVEs as relevant, for which a taxonomy backfill (repeat Phase 4) is then performed.
\subsubsection{Phase~6 - Security Assessment} In Phase 6, the SAR is generated based on the enriched KG, the relevant CVEs, and the constructed attack paths.
The taxonomy chains are extracted by traversing the KG and incorporated as evidence in each CVE observation. In addition, the CRH, the EOL status, and the affected SSP component are included in the observations. The EOL context enables the agent to explicitly highlight affected assets without an available patch path as a strategic risk in the report.
Both SSP and SAR are then validated against the NIST OSCAL JSON schemas. By passing the strict schema validation without errors, the artifacts confirm audit-readiness for direct integration into existing audit workflows. Altogether, the pipeline produces five output artifacts: OSCAL SSP, OSCAL SAR, CRH-based Risk Assessment (JSON), KG (JSON), and BSI-Grundschutz++ Gap-Analysis Report (JSON).
\subsection{Dataset - Evidence-Based Reference Architecture}
The reference scenario WaterWork represents a regional water facility with high-pressure pumps, chlorine dosing, and SCADA control. The architecture is modeled after documented patterns from real incidents~\cite{wds_architecture}. The architecture includes a Cisco ASA~5505 (Firmware v9.2.4) as an internet-exposed VPN gateway without multi-factor authentication, an engineering workstation running Windows~7~SP1 with Advantech WebAccess/SCADA~v8.2, a historian server based on Microsoft SQL Server~2012 Express (Windows Server~2012~R2), and six Siemens SIMATIC S7-1200 CPUs (Firmware v4.0) communicating through S7Comm on port~102 directly from the SCADA workstation. A vendor laptop from WaterTech Solutions with TeamViewer~11 is permanently connected to the network. There is no firewall segmentation between IT and operational technology zones (flat network on 10.10.10.0/24). The pipeline input consists of a Shodan report that exposes a single public IP with two services (TCP/443 AnyConnect, UDP/500 ISAKMP), the Cisco ASA CPE string, and CVE-2018-0101. Since the evaluation operates without live infrastructure, Shodan reports are provided as textual input to the pipeline. In a production deployment, the data stream would be obtained via the MCP-Shodan interface.
In order to allow for a quantitative evaluation, we created a manually verified ground truth for the WaterWork scenario by listing the entire CVE surface area for each component through cross-referencing NVD CPE match criteria, vendor advisories, and ICS-CERT publications, yielding 292~CVEs across eight assets.  Each CVE is assigned a validity rating for the respective version, a CVSS base score, and the current EPSS probability. The verified CVE set consists of 15 multi-stage causal attack paths that link perimeter breaches, lateral movements, and operational technology impacts, using only CVEs that have been confirmed for the versions used. Additionally, the ground truth includes 16~ATT\&CK Enterprise technique mappings and 34~D3FEND countermeasure mappings with defense strategies for each attack path. This ground truth serves as a reference basis for measuring pipeline recall, triage decision accuracy, and actual hallucination rate.
\subsection{Evaluation Design}
The WaterWork scenario was assessed with the eight-phase pipeline. OpenAI GPT-4.1~\cite{openai_model} served as the base LLM. The deterministic phases (1, 2, 3a, 4) could access authoritative data sources through 15~active MCP servers, as illustrated in Table~\ref{tab:mcp_servers}. The scenario was executed in five independent runs with identical input to quantify the variance of the agent-based phases, and the pipeline outputs are evaluated against the ground truth.
\section{Results \& Discussion}
Across five independent runs, the pipeline consistently identified 8~entities and discovered $398 \pm 9$~CVEs.
For each node type in the KG, we verify its existence against the authoritative source. The hallucination rate is the fraction of nodes that fail this check. All CVE nodes in the KG ($380 \pm 9$ per run) correspond to valid NVD entries.
As expected, the Factual Hallucination Rate for all KG nodes is 0\%, meaning that no hallucinated identifiers were observed. While the deterministic MCP architecture guarantees a Factual Hallucination Rate of 0\% (i.e., no fictitious CVEs or synthetic CVSS scores are generated), the pipeline remains susceptible to Contextual Hallucinations originating from Phase~0. If the LLM incorrectly extracts an asset from the unstructured text (e.g., a generic "Windows" entity when only Windows~7~SP1 and Windows Server~2012~R2 are described), the deterministic retrieval logically propagates this error, retrieving valid CVEs for an absent asset. We define the Semantic Hallucination Rate in~\eqref{eq:shr} where entity classifications are determined against the ground truth.
\begin{equation}
\text{SHR} = \text{FP}_{\text{entities}} / (\text{TP}_{\text{entities}} + \text{FP}_{\text{entities}})
\label{eq:shr} 
\end{equation}
Precision and Recall are defined as:
\begin{align}
 \text{Recall} &= TP \,/\, (TP + FN) \label{eq:recall}\\
 \text{Precision} &= TP \,/\, (TP + FP) \label{eq:precision}
 \end{align} 
Across all five runs, Phase~0 achieved an Entity Extraction Precision of 87.5\% and a Recall of 100\%, with an Semantic Hallucination Rate of 12.5\% (1~FP entity, and 7/8 correct entities per run). Version extraction accuracy is 100\% (7/7 correct across all runs).
The single FP entity ("Windows (unknown version)") was assigned the broadest possible CPE, causing the deterministic Phase~1 to retrieve 55~CVEs, of which 50 are unique to this entity. Of these 50~unique CVEs, 30 survive triage and appear in the final SAR, yielding a Contextual False Positive Rate of $8.5 \pm 0.2\%$ (30 out of $350 \pm 9$ SAR CVEs). The 30~Contextual False Positives are real CVEs for a non-existent generic Windows asset that pass the noise filter.
Traditional RAG often fails at generation time by fabricating identifiers or scores. MCP-bounded retrieval removes that failure mode, but it introduces a different one, as a single semantic error in Phase~0 can propagate deterministically and still contribute irrelevant CVEs to the final SAR. In contrast to RAG’s distributed stochastic errors, the hallucination range shifts upstream to entity extraction phase and its precision. This concentrates the error surface in one identifiable phase, making failures both higher-impact and easier to localize. Consequently, the reliability of the pipeline is bounded by the precision of Phase~0, rendering entity extraction the critical control point for system reliability.

The five independent runs confirm that the deterministic phases produce identical outputs for identical inputs. Entity extraction (Phase~0) is extracting the same 8~entities in every run across all five runs. The CVE set stability is high, as the core intersection across all five runs comprises 352~CVEs and the union 397~CVEs, with a mean pairwise Jaccard of $0.95 \pm 0.05$. The variance is attributable exclusively to LLM phases, as Phase~3b (triage rehabilitation) invoked $2.2 \pm 1.6$~tool calls per run, while Phase~5 (LLM promotion) consistently invoked 0~tool calls, indicating no additional CVEs were promoted by the agent in this scenario. 
Against 292 ground truth CVEs, the pipeline achieves a CVE Recall of $0.90$ and a Precision of $0.74$ across five runs, with 263 of 292 ground truth-CVEs matched per run. Of the 29~consistently missed CVEs, 26~are Adobe Flash/Reader CVEs and Windows Server~2012~R2 font-rendering CVEs discovered via the broad Windows~OS CPE. These are correctly filtered by Phase~5 as irrelevant to the operational technology deployment context. Only one genuine miss remains, CVE-2017-14016 (Advantech WebAccess), which was found neither through CPE search nor via ICS-CERT advisory search, indicating a coverage gap for older ICS advisories. 
Against the expanded ground truth of 15~ATT\&CK techniques, the pipeline achieves a Recall of $0.94$ and a Precision of $0.90 \pm 0.03$, matching 14 of 15 techniques per run.
Taxonomy chain completeness is calculated as follows, where $c_i \in \{0,\ldots,4\}$ counts populated levels per CVE:
\begin{align} 
\bar{C} = \frac{1}{n}\sum c_i
\end{align} 
The pipeline achieves a D3FEND Recall of $1.00$ and a Precision of $0.88 \pm 0.01$, matching all $34$~ground truth countermeasures per run. The completeness of taxonomy chains reaches $0.16$ vs.\ $2.80$ in ground truth, because only the top-25 high-urgency CVEs undergo a full chain expansion.
To illustrate the qualitative output beyond aggregate metrics, the KG traces a multi-stage path from the internet-exposed Cisco ASA~5505 (CVE-2016-6366, CWE-120 Buffer Overflow $\to$ CAPEC-100 $\to$ T1068 Exploitation for Privilege Escalation $\to$ D3-MBT Memory Boundary Tracking) through the unpatched Windows~7~SP1 workstation (CVE-2019-0708, BlueKeep) to the SQL Server historian (CVE-2020-0618, CWE-502 Deserialization of Untrusted Data $\to$ CAPEC-586 $\to$ T1059 Command and Scripting Interpreter $\to$ D3-CF Content Filtering), with each node verified against its authoritative source.

The CRH prioritization is compared against SSVC, which encodes practitioner consensus on vulnerability prioritization~\cite{ssvc_score}. For the 318 CVEs with CVSS v3.x vectors, Spearman rank correlation (a non-parametric measure of ranking agreement, where $\rho = 1$ indicates identical ordering) between CRH and SSVC is $\rho = 0.797$, exceeding the CVSS-SSVC baseline of $\rho = 0.677$. The improvement reflects the operational technology context factors (asset criticality, exposure) that CRH adds beyond raw CVSS. The near-agreement rate (exact or $\pm 1$ ordinal class) is 97.5\% (310/318). Since both metrics share CVSS, EPSS, and KEV as inputs, this concordance should be read as an indirect consistency check rather than independent validation.
To ensure CRH robustness, a one-at-a-time sensitivity analysis varies each CRH parameter independently while holding others at baseline values. Four parameters are swept: SCADA floor~$S_{\text{base}}$, exposure multiplier~$E_{\text{exp}}$, KEV penalty, and triage threshold~$U$. For each, triage volume, class migrations, and ground truth retention are reported. The full Phase~3a triage logic applies three OR-rules: urgency above absolute threshold, CISA KEV listed, and urgency above role threshold~$>5$ for Gateway/Target assets. All ground truth CVEs in the triage-relevant set are retained across all~14 configurations. The triage threshold is the dominant sensitivity parameter (triage volume range = 217-330~CVEs), while SCADA floor, exposure multiplier, and KEV penalty have zero effect on triage volume under the role-based safety rule. Notably, CVE-2019-10943 (Siemens S7-1200, urgency~$11.26$) falls below the absolute threshold of~$12.00$ but is retained via the role-based rule (Target asset, urgency~$> 5.00$), demonstrating that the multi-rule triage design provides structural robustness against parameter variation.
For a safety-first heuristic in critical infrastructure environments, the relevant validation criterion is decision stability, as no critical CVE must be lost under moderate parameter variation. The 100\% ground truth retention across all 14~configurations directly evidences this property. The role-based safety rule (Gateway/Target $> 5$) creates a structural safety net that decouples triage outcomes from the choice of individual parameters, even when the absolute threshold varies by $\pm 50$\%; the role-based rule independently retains all CVEs affecting exposed or target assets. This multi-rule design renders the system robust by construction against moderate parameter variations, rather than requiring precise calibration of any single parameter.
The OSCAL export (SSP and SAR) passes strict NIST OSCAL v1.1.2 JSON Schema validation without errors across all five runs, confirming both the structural correctness of the internal KG and audit-readiness for direct integration into, e.g., the Grundschutz++ automated continuous compliance workflow.
\section{Limitations \& Future Work} 
Due to the absence of publicly available critical infrastructure or OSCAL datasets, the evaluation relies on only a single synthetic evidence-based reference architecture, so cross-scenario variance remains unquantified. The CRH heuristic is transparent but still needs to be evaluated further, and the SSVC comparison should be read as indirect validation rather than as a substitute for a dedicated expert study. The pipeline also remains limited to software-vulnerability-centric attack paths and does not model threats such as credential theft, insider abuse, or social engineering. The Phase~0 entity extraction constitutes the single LLM-dependent trust boundary in the pipeline. As demonstrated by the 12.5\% Semantic Hallucination Rate and resulting 8.5\% Contextual False Positive Rate, errors at this boundary propagate deterministically through all downstream phases. A human-in-the-loop review after Phase~0 could interrupt this cascade. As legacy operational technology systems rarely change, an analyst reviewing the extracted entities before the deterministic pipeline begins would break the error cascade at its origin.
Furthermore, the evaluation depends on the timeliness and coverage of curated and periodically updated CTI sources. Therefore, temporal publication lag in CTI data or gaps in advisory feeds can affect triage accuracy. An advantage of the pipeline is its suitability for continuous assessment. Due to the low cost per run, the pipeline can be operated iteratively on a continuous basis. This mitigates temporal gaps in CTI source coverage.
Finally, the BSI Grundschutz++~\cite{bsi_sdtb_github} is currently a preview and is being incrementally expanded. The finalized version is expected by the end of 2026. Until then, the catalog has no formal certification status, and Grundschutz++-based assessments are informational only. However, as the pipeline fetches the latest version from the BSI GitHub repository at runtime~\cite{bsi_sdtb_github}, rerunning the pipeline mitigates this limitation.
Future work should include additional evidence-based scenarios from energy, transport, and other critical infrastructure sectors to test cross-domain robustness, and a structured concordance study with operational technology practitioners to provide stronger external validation of the CRH heuristic.
\section{Conclusion}
This paper presented a proof-of-concept for an MCP-grounded eight-phase agent pipeline for converting unstructured critical infrastructure documentation into audit-ready OSCAL security artifacts without active scanning of/within the critical infrastructure. 
The main contribution is a deterministic bounded-trust architecture in which MCP retrieval prevents fabricated identifiers in the KG, while the evaluation shows that residual failure is dominated by upstream semantic imprecision rather than downstream fabrication.
In the WaterWork reference scenario, the system achieves strong recall of vulnerabilities and defensive mappings, produces valid SSP and SAR artifacts, and exposes a clear remaining weakness in entity extraction. For high-stakes operational technology workflows, this is a useful trade-off, as the error surface becomes narrower, more observable, and easier to control. The next step is, therefore, not more downstream generation, but better trust-boundary management at the point where raw language becomes structured assets.

\end{document}